\documentclass[aps,preprint]{revtex4}%
\usepackage{amsfonts}
\usepackage{amsmath}
\usepackage{amssymb}
\usepackage{graphicx}%
\providecommand{\U}[1]{\protect\rule{.1in}{.1in}}

\begin{document}
\preprint{ }
\title[Inhomogeneous string]{The vibrating inhomogeneous string: a topic for a course in Computational Physics}
\author{George Rawitscher}
\affiliation{Physics Department, University of Connecticut, Storrs CT}
\author{Jakob Liss}
\affiliation{International Fulbright exchange student, Physics Department, University of
Connecticut, Storrs CT}

\begin{abstract}
This paper solves the integral equation which describes the oscillating
inhomogeneous string, by using a spectral expansion method in terms \ of
Chebyshev polynomials. The result is compared with the solution of the
corresponding differential equation, obtained by expansion into a set of
sine-wave functions, with emphasis on the accuracies of the two methods. These
accuracies are determined by comparison with an iterative method which allows
a precision of $1:10^{11}$. The iterative method is based on a old method by
Hartree, but contains innovative spectral expansion procedures. \TeX{}

\end{abstract}
\startpage{1}
\maketitle
\tableofcontents

\section{Introduction}

The teaching of computational physics courses is now practiced by many
universities, and excellent text books are available supporting this endeavor
\cite{BOOKS}, as well as papers describing such courses \ \cite{PAPERS}. In
particular, the vibrating string provides an excellent topic \cite{STRING},
since the solution of the corresponding differential equation can be achieved
by several different ways, and useful comparison between the different methods
can be provided.

If the string is inhomogeneous, the separation of variables method becomes
more involved than for the homogenous case, since the spatial part becomes the
solution of a Sturm-Liouville (SL) eigenvalue equation and is no longer a
simple sine wave. The SL equation is usually solved by expansion into a basis
set of functions (sine waves for the clamped string) that will lead to a
matrix equation for the expansion coefficients. The eigenvalues and
eigenvectors of this matrix then provide the SL functions, but, the accuracy
depends on the size of the basis, and correspondingly on the size of the
matrix. The accuracy of this method can be studied by introducing an entirely
different method of solution of the SL equation, which is normally not
discussed in the existing teaching literature. This method, denoted as $IEM$
(for Integral Equation Method)$,$ consists in transforming the SL differential
equation into an equivalent integral equation, and solving the latter by an
expansion into Chebyshev polynomials \cite{IEM}, \cite{CISE}. This method has
the advantage that its accuracy can be automatically pre-determined by means
of an accuracy parameter, the number of mesh points required to achieve a
particular accuracy is much smaller than for the more conventional finite
difference methods (by a factor close to 20), and the size of the matrices is
kept small by a partition technique, thus avoiding the drawbacks of large
matrices in conventional integral equation solution methods. These advantages
are important for the solution of a computationally complex problem
\cite{COMPLEX}. It is the purpose of the present paper to explain the $IEM$
method in simple terms, and apply it to the solution of the inhomogeneous
vibrating string. A method to solve the SL iteratively, thus avoiding the
introduction of the inaccuracies described above, will also be presented. This
method was first devised by Hartree \cite{HART} in the solution of atomic
physics energy eigenvalues of the Schr\"{o}dinger equation. It has now been
adapted to the spectral $IEM$ solution of the equivalent integral equation
\cite{HEHE}, and since it can achieve an accuracy of $1:10^{11}$ it does
provide the bench mark values against which the previous methods for the
inhomogeneous string can be compared.

The method for the solution of the string equation is very close to the
solution of the important quantum mechanical time independent Schr\"{o}dinger
equation. Since the properties of the string are much easier to visualize than
the properties of the Schr\"{o}dinger equation, the present discussion of the
vibrations of the string also serves as a pedagogical introduction to the
numerical methods required for quantum mechanics. The numerical calculations
are done with MATLAB. An excellent introduction into both MATLAB\ and
numerical methods can be found in the book by Recktenwald \cite{RECK}. The
MATLAB programs for the calculations presented here will be available in the
"compadre" digital library \cite{PADRE}.

In summary, the main purpose of this paper is to introduce to the teaching
community the use of spectral expansions, especially for the solution of
integral equations, because of its elegance, its accuracy, and its
computational economy. The method of spectral expansions is not new (since
circa 1970) and is described in the excellent book by L. N. Trefethen
\cite{SPECTRAL}.

Section $2$ presents the differential equations describing the vibrating
inhomogeneous string and the solution by expansion into a basis of sine-wave
functions; Section $3$ presents the basics of expansions into Chebyshev
polynomials, section $4$ presents the Sturm-Liouville ($SL$) integral equation
that is equivalent to the differential equation, and also presents the
solutions in terms of the $IEM$ \ spectral method; in section $5$ the
iterative solution of the $SL$ equation is described, and the accuracy of the
previous methods is examined. Section $6$ contains a summary and conclusions.

\section{The inhomogeneous vibrating string}

Consider a stretched string of metal, clamped between two horizontal points
$P_{1}$ and $P_{2}$. The distance between the fixed points is $L,$ the mass
per unit length $\rho$\ of the string is not a constant, as described below,
and the speed of propagation of the waves depends on the location along the
string. When a disturbance is excited along the string, the particles on the
string vibrate in the vertical direction with a distribution of frequencies to
be determined.

Denote by $y(x,t)$ the (small) displacement of a point on the string in the
vertical direction away from the equilibrium position $y=0$, for a given
horizontal distance $x$ of the point from the left end $P_{1}$, and at a time
$t.$ As can be shown, the wave equation is
\begin{equation}
\frac{\partial^{2}y}{\partial x^{2}}-\frac{\rho}{T}\frac{\partial^{2}%
y}{\partial t^{2}}=0 \label{9-1}%
\end{equation}
where $T$ is the tension along the string. We define a function $R(x)$ which
is dimensionless, and which describes the variation of $\rho$ with $x$
according to
\begin{equation}
\rho(x)=\rho_{0}\ R(x) \label{9-2}%
\end{equation}
where $\rho_{0}$ is some fixed value of $\rho.$\ Defining a reference speed
$c$ according to
\begin{equation}
\frac{\rho_{0}}{T}=\frac{1}{c^{2}} \label{9-3}%
\end{equation}
the wave equation becomes
\begin{equation}
\frac{\partial^{2}y}{\partial x^{2}}-\frac{1}{c^{2}}R(x)\frac{\partial^{2}%
y}{\partial t^{2}}=0 \label{9-4}%
\end{equation}

According to the solution by means of separation of variables, $y(x,t)=\psi
(x)\ A(t),$ one obtains the separate equations
\begin{equation}
\frac{d^{2}\psi(x)}{dx^{2}}+\Lambda R(x)\ \psi(x)=0 \label{9-5}%
\end{equation}
and%
\begin{equation}
\frac{d^{\ 2}A}{dt^{2}}=-\Lambda c^{2}A \label{9-6}%
\end{equation}
We assume that the constant $\Lambda$ is positive. A general solution for Eq.
(\ref{9-6}) is $a\cos(wt)+b\sin(wt),$ with
\begin{equation}
w=c\sqrt{\Lambda} \label{9-7}%
\end{equation}
where $\Lambda$ is an eigenvalue of Eq. (\ref{9-5}).

The Eq. (\ref{9-5}) is a Sturm-Liouville equation \cite{BOAS} with an infinite
set of eigenvalues $\Lambda_{n},\ \ n=1,2,3,...$ and the corresponding
eigenfunctions $\psi_{n}(x)$ form a complete set, denoted as "sturmians", in
terms of which the general solution can be expanded
\begin{equation}
y(x,t)=\sum_{n=1}^{\infty}[a_{n}\ \cos(\omega_{n}t)+b_{n}\ \sin(\omega
_{n}t)]\ \psi_{n}(x) \label{9-8}%
\end{equation}
where $w_{n}=c\sqrt{\Lambda_{n}}.$ The objective is to calculate the functions
$\psi_{n}(x)$ and the respective eigenvalues $\Lambda_{n}$ as solutions of Eq.
(\ref{9-5}), with the boundary conditions that $y=0$ for $x=0$ and $x=L$,
\begin{equation}
\psi_{n}(0)=\psi_{n}(L)=0, \label{9-8bound}%
\end{equation}
and that for $t=0$
\begin{equation}
y(x,0)=f(x)\text{ and \ }dy/dt|_{t=0}=g(x). \label{9-9bound}%
\end{equation}
The constants $a_{n}$ and $b_{n}$ in Eq. (\ref{9-8}) are obtained from the
initial displacement of the string from its equilibrium position $f(x)$ and
$g(x)$, in terms of integrals of that displacement over the functions
$\psi_{n}(x).$%
\begin{equation}
a_{n}=\int_{0}^{L}f(x)\ \psi_{n}(x)\ dx;~~b_{n}=\frac{1}{\omega_{n}}\int
_{0}^{L}g(x)\ \psi_{n}(x)\ dx. \label{9-10}%
\end{equation}

\subsection{The case of the homogeneous string.}

In the case that the string is homogeneous, the function $R(x)=1$ becomes a
constant, and the Sturmian functions are given by the sine functions, i.e.,
$\psi_{n}(x)=\phi_{n}(x)$, with%
\begin{equation}
\phi_{n}(x)=\sqrt{2/L}\ \sin(k_{n}x)\text{, ~~}k_{n}=n(\pi/L),~~\text{\ }%
n=1,2,3... \label{9-9}%
\end{equation}
and the eigenvalues become $\Lambda_{n}=k_{n}^{2}=[n\pi/L]^{2}.$ Assuming that
the initial displacement functions $f(x)$ and $g(x)$ of the string are given
by
\begin{equation}
\ f(x)=x\sin[(\pi/L)x],\ g(x)=0 \label{9-11bound}%
\end{equation}
and
\begin{equation}
L=1m,\ c=800\ m/s. \label{9-12bound}%
\end{equation}
then one can evaluate Eq. (\ref{9-10}) for the coefficients $a_{n}$
analytically (all the $b_{n}=0).$ One finds that all $a_{n}$ vanish for $n$
odd, with the exception for $n=1$, for which%
\begin{equation}
a_{1}=-\frac{L^{2}}{4}\sqrt{\frac{2}{L}} \label{II-9}%
\end{equation}
For $n$ even, the corresponding result for $a_{n}$ is
\begin{equation}
a_{n}=\frac{L^{2}}{\pi^{2}}\sqrt{\frac{2}{L}}\ \left[  \frac{1}{(1+n)^{2}%
}-\frac{1}{(1-n)^{2}}\right]  ,~~n=2,4,... \label{II-8}%
\end{equation}
With the above results the truncated sum (\ref{9-8})
\begin{equation}
y^{(n\max)}(x,t)=\sum_{n=1}^{n\max}[a_{n}\ \cos(\omega_{n}t)+b_{n}%
\ \sin(\omega_{n}t)]\ \phi_{n}(x) \label{II-10}%
\end{equation}
can be calculated. The result is displayed in Figs. (\ref{FIGII1}) and
(\ref{FIGII2})
\begin{figure}
[ptb]
\begin{center}
\includegraphics[
trim=-0.060228in 0.030141in 0.060227in -0.030141in,
height=2.0418in,
width=2.7146in
]%
{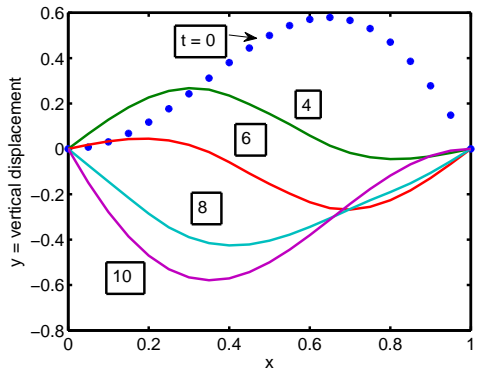}%
\caption{Vibrations on the homogeneous string. The symbols * mark the initial
displacement of the string from its equilibrium position, given by
Eq.(\ref{9-11bound}). The numbers written next to each curve indicate the
time, in units of $L/c$}%
\label{FIGII1}%
\end{center}
\end{figure}
\begin{figure}
[ptb]
\begin{center}
\includegraphics[
height=2.098in,
width=2.7908in
]%
{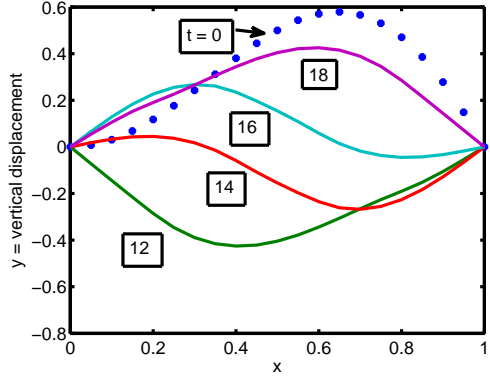}%
\caption{Continuation form Fig.(\ref{FIGII1}) of the time development of the
vibrations of the string. }%
\label{FIGII2}%
\end{center}
\end{figure}

For $n\gg1$, $a_{n}$ will approach $0$ like $(1/n)^{3},$ i.e., quite slowly.
It is desirable to examine how many terms are needed in the numerical sum of
Eq. (\ref{II-10}) in order to get an accuracy of $4$ significant figures in
$y.$ A good guess is that the sum of all terms not included in the sum
\begin{equation}
\sum_{n\max+1}^{\infty}a_{n}\cos(\omega_{n}t)\simeq-4\frac{L^{2}}{\pi^{2}%
}\sqrt{\frac{2}{L}}\int_{n\max+1}^{\infty}\frac{1}{n^{3}}\cos(\frac{c\pi}%
{L}t\ n)\ dn \label{II11}%
\end{equation}
should be less than $y_{\max}\times10^{-4}.$ The integral in Eq. (\ref{II11})
is smaller than $\int_{n\max+1}^{\infty}(1/n)^{3}dn=(n_{\max}+1)^{-2}/2$
(since the $\cos$ term produces cancellations), and one obtains the estimate%
\begin{equation}
|\sum_{n\max+1}^{\infty}a_{n}\cos(\omega_{n}t)|\ <2\frac{L^{2}}{\pi^{2}}%
\sqrt{\frac{2}{L}}(n\max+1)^{-2} \label{II12}%
\end{equation}
With $nmax=50$ the right hand side of Eq. (\ref{II12}) is $\simeq10^{-4}.$ A
numerical evaluation of the difference $|y^{(50)}(x,0)-f(x)|\ $is less
than$\ 10^{-5},$ which confirms that with $n\max=50$ the accuracy expected for
$y^{(50)}(x,t)$ is better than $1:10^{4}.$

\subsection{The inhomogeneous string by means of a Fourier series expansion}

An approximate solution to Eq. (\ref{9-5}) for $\psi_{n}$ is to expand it in
terms of the Fourier sine waves given by Eq. (\ref{9-9}), since these
functions obey the same boundary conditions as the $\psi_{n}^{\prime}s.$ The
approximation consists in truncating that expansion at an upper limit
$\ell\max=N,$ and also drop the sub- and -superscript $(n)$ for the time being%
\begin{equation}
\psi^{(N)}(x)=\sum_{\ell^{\prime}=1}^{N}d_{\ell^{\prime}}\phi_{\ell^{\prime}%
}(x). \label{9-12}%
\end{equation}
Inserting expansion (\ref{9-12}) into Eq. (\ref{9-5}), remembering that
$d^{2}\phi_{\ell}(x)/dt^{2}=-k_{\ell}^{2}\phi_{\ell}(x)$, multiplying Eq.
(\ref{9-5}) by a particular function $\phi_{\ell}(x)$, integrating both sides
of the equation over $dx$ from $x=0$ to $x=L,$ and using the orthonormality of
the functions $\phi_{\ell}(x),$ one obtains%
\begin{equation}
-k_{\ell}^{2}\ d_{\ell}+\Lambda\sum_{\ell^{\prime}=1}^{N}R_{\ell,\ell^{\prime
}}d_{\ell^{\prime}}=0 \label{9-13}%
\end{equation}
where
\begin{equation}
R_{\ell,\ell^{\prime}}=\int_{0}^{L}\phi_{\ell}(x)R(x)\phi_{\ell^{\prime}%
}(x)\ dx \label{9-14}%
\end{equation}
are the matrix elements of the function $R$ over the basis functions
$\phi_{\ell}.$ This equation (\ref{9-13}) can also be written in matrix form,
where
\begin{equation}%
\begin{pmatrix}
k_{1}^{2} &  &  &  & \\
& k_{2}^{2} &  &  & \\
&  & k_{3}^{2} &  & \\
&  &  & \ddots & \\
&  &  &  & k_{N}^{2}%
\end{pmatrix}%
\begin{pmatrix}
d_{1}\\
d_{2}\\
d_{3}\\
\vdots\\
d_{N}%
\end{pmatrix}
=\Lambda%
\begin{pmatrix}
R_{1,1} & R_{1,2} & R_{1,3} & \cdots & R_{1,N}\\
R_{2,1} & R_{2,2} & R_{2,3} & \cdots & R_{2,N}\\
R_{3,1} & R_{3,2} & R_{3,3} & \cdots & R_{3,N}\\
\vdots & \vdots & \vdots & \ddots & \vdots\\
R_{N,1} & R_{N,2} & R_{N,3} & \cdots & R_{N,N}%
\end{pmatrix}%
\begin{pmatrix}
d_{1}\\
d_{2}\\
d_{3}\\
\vdots\\
d_{N}%
\end{pmatrix}
, \label{9-16}%
\end{equation}
or more succinctly%
\begin{equation}
\mathbf{k}^{2}\vec{d}=\Lambda\ \mathbf{R}\vec{d}\mathbf{,} \label{9-15}%
\end{equation}
where bold letters indicate matrices, and a vector quantity indicates a
$(N\times1)$ column. Since all the $k_{\ell}$'s are positive, the matrix
$\mathbf{k}^{-1}$ can be defined as%
\begin{equation}
\mathbf{k}^{-1}=%
\begin{pmatrix}
k_{1}^{-1} &  &  &  & \\
& k_{2}^{-1} &  &  & \\
&  & k_{3}^{-1} &  & \\
&  &  & \ddots & \\
&  &  &  & k_{N}^{-1}%
\end{pmatrix}
\label{9-17}%
\end{equation}
and one can transform Eq. (\ref{9-15}) into%
\begin{equation}
\mathbf{M}_{fourier}\mathbf{\ }\vec{u}_{n}\mathbf{=}\frac{1}{\Lambda_{s}}%
\vec{u}_{n};~~~n=1,2,...N \label{9-18}%
\end{equation}
where
\begin{equation}
\mathbf{M}_{fourier}\mathbf{=k}^{-1}\mathbf{R\ k}^{-1} \label{9-19}%
\end{equation}
and%
\begin{equation}
\vec{u}_{n}\ \mathbf{=k\ }\vec{d}_{n}\ \mathbf{.} \label{9-20}%
\end{equation}
While Eq. (\ref{9-15}) is a generalized eigenvalue equation, Eq. (\ref{9-18})
is a simple eigenvalue equation. The vectors $\vec{u}_{n}$ are the $N$
eigenvectors of the \ $N\times N$\ matrix $\mathbf{M}_{fourier}$, and
$1/\Lambda_{n}$ are the eigenvalues. Furthermore, since $\mathbf{R}$ is a
symmetric matrix, $\mathbf{M}_{fourier}$ is also symmetric. The eigenvectors
of a symmetric matrix are orthogonal to each other, i.e. $(\vec{u}_{n}%
)^{T}\cdot\vec{u}_{m}=\delta_{n,m}.$ Here $T$ indicates transposition. However
the vectors $\vec{d}_{n}$ are not orthogonal to each other, since $(\vec
{d}_{n})^{T}\cdot\vec{d}_{m}=(\vec{u}_{n})^{T}\mathbf{k}^{-2}\vec{u}_{m}.$

In summary, the procedure is as follows

1. Choose an upper truncation limit $N$ of the sum (\ref{9-12});

2. Calculate the matrix elements $R_{\ell,\ell^{\prime}}$ so as to obtain the
$N\times N$ matrix $\mathbf{R}$

3. Construct the matrix $\mathbf{M}_{fourier}$ from Eq. (\ref{9-19}), and find
the eigenvalues $(1/\Lambda_{n})$ and eigenvectors $\vec{u}_{n}$, $n=1,2,..N,$
by using the MATLAB eigenvalue command $[\mathbf{V},\mathbf{D}]=eig(\mathbf{M}%
).$ The output $\mathbf{D}$ is a diagonal matrix of the eigenvalues and
$\mathbf{V}$ is a full matrix whose columns are the corresponding eigenvectors
so that $\mathbf{M}\ast\mathbf{V}$ = $\mathbf{V}\ast\mathbf{D}$. For example
$\vec{u}_{n}=\mathbf{V}(:,n)$

4. If $\vec{\Phi}(x)$ is the column vector of the $N$ basis functions
$\phi_{\ell}(x)$, then $\psi(x)$ can be written as (the superscript $(N)$ is
dropped now)
\begin{equation}
\psi_{n}(x)=(\vec{u}_{n})^{T}\mathbf{k}^{-1}\cdot\vec{\Phi}(x). \label{9-21}%
\end{equation}

5. In view of Eq. (\ref{9-21}) the coefficients $a_{n}=$and $b_{n}$ $=<g$
$\psi_{n}>$ can be written as
\begin{align}
a_{n}  &  =<f\psi_{n}>\ =(\vec{u}_{n})^{T}\mathbf{k}^{-1}\cdot<f\ \vec{\Phi
}(x)>\label{9-22a}\\
b_{n}  &  =<g\psi_{n}>\ =(\vec{u}_{n})^{T}\mathbf{k}^{-1}\cdot<g\ \vec{\Phi
}(x)> \label{9-22b}%
\end{align}
\newline where $\langle f\ \vec{\Phi}(x)\rangle$ is the column vector of the
integrals $\langle f\ \phi_{\ell}\rangle=\int_{0}^{L}f(x)\phi_{\ell}(x)dx,$
$\ell=1,2,..N.$

6. The final expression for $y(x,t)$ can be obtained by first obtaining the
coefficients $e_{n}$%
\begin{equation}
e_{n}(t)=(\vec{u}_{n})^{T}\mathbf{k}^{-1}\left[  \langle f\ \vec{\Phi
}(x)\rangle\cos(w_{n}t)+\langle g\ \vec{\Phi}(x)\rangle\frac{1}{w_{n}}%
\sin(w_{n}t)\right]  , \label{9-23}%
\end{equation}
and then performing the sum%
\begin{equation}
y(x,t)=\sum_{n=1}^{N}e_{n}(t)\psi_{n}(x)=\vec{e}^{\ T}\cdot\vec{\Psi}.
\label{9-24}%
\end{equation}
In the above, $\vec{e}$ is the column vector of all $e_{n}$'s, and $\vec{\Psi
}$ is the column vector of all $\psi_{n}$'s. In the present discussion we
limit ourselves to calculating the eigenvalues $\Lambda_{n}.$

Assuming that the mass per unit length changes with distance $x$ from the left
end of the string as
\begin{equation}
R(x)=1+2x^{2}, \label{9-25}%
\end{equation}
and $c,$ $f(x)$ and $g(x)$ are the same as for the homogeneous string case,%
\begin{equation}
L=1m,\ c=800\ m/s,\ f(x)=x\sin[(\pi/L)x],\ g(x)=0. \label{9-26}%
\end{equation}
then the integrals (\ref{9-14}) for the matrix elements $R_{\ell,\ell^{\prime
}}$ can be obtained analytically with the result%
\begin{align}
R_{\ell,\ell^{\prime}}  &  =2\ast2\left(  \frac{L}{\pi}\right)  ^{2}%
(-1)^{\ell+\ell^{\prime}}\left[  \frac{1}{(\ell-\ell^{\prime})^{2}}-\frac
{1}{(\ell+\ell^{\prime})^{2}}\right]  ,\ \ \ell\neq\ell^{\prime}%
\label{9-27a}\\
R_{\ell,\ell}  &  =1+2\ L^{2}\left[  \frac{1}{3}-\frac{2}{(2\pi\ell)^{2}%
}\right]  ,\ \ \ \ell=\ell^{\prime}. \label{9-27b}%
\end{align}
The increase of $R$ with $x$ can be simply visualized with the choice
(\ref{9-25}). More realistic situations, such as the distribution of masses on
a bridge, can be envisaged for future applications.

The numerical construction of the matrices $\mathbf{R}$\textbf{ }and\textbf{
}$\mathbf{M}_{fourier}$ is accomplished in the MATLAB program
$string\_fourier.m$ which in turn calls the function $inh\_str\_M.m$, using
the input values%
\begin{equation}
L=1m,~~c=800\ m/s,\ \label{9s-4}%
\end{equation}
The truncation value $N$ of the sum Eq. (\ref{9-12}) is set equal to either
$30$ or $60$, and the corresponding dimension of the matrices $\mathbf{M}%
_{fourier}$ or $\mathbf{R}$ is $N\times N.$ These values are chosen so as to
examine the sensitivity of the eigenvalues to the size of the matrix
$\mathbf{M}_{fourier}\mathbf{.}$

\begin{figure}
[ptb]
\begin{center}
\includegraphics[
height=2.5374in,
width=3.3745in
]%
{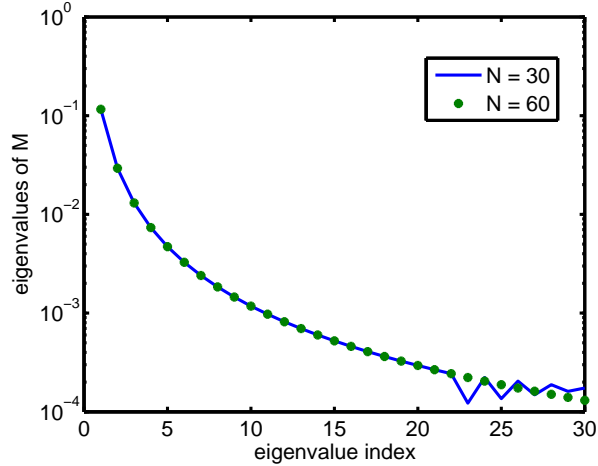}%
\caption{The eigenvalues of the matrix $M_{fourier},$ defined in
Eq.(\ref{9-19}). \textbf{\ }The quantity $N$ indicates the truncation value of
the sum in Eq. (\ref{9-12}), that expands the string displacement
eigenfunction $\psi_{n}(x)$ into the Fourier functions $\phi_{\ell}(x)$. The
dimension of the matrix $M_{fourier}$ is $N\times N$. }%
\label{FIG9s_1}%
\end{center}
\end{figure}

\begin{figure}
[ptb]
\begin{center}
\includegraphics[
height=2.501in,
width=3.3252in
]%
{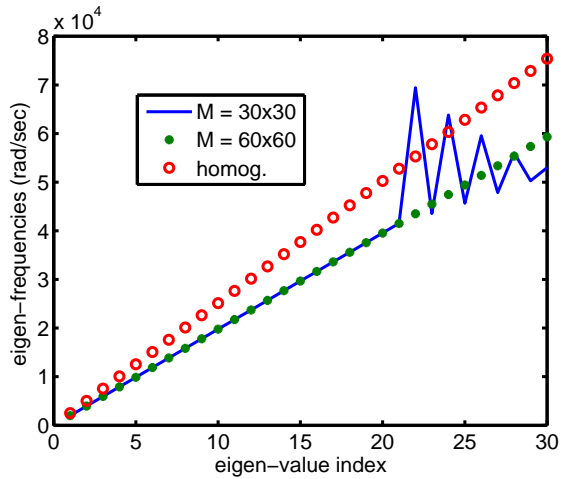}%
\caption{The frequencies in units of radians/sec of the vibration of the
inhomogeneous string, compared with the frequencies of the corresponding
homogeneous string. The higher frequencies become inaccurate when the
dimension of the matrix $M_{fourier}$ is too small.}%
\label{FIG9s_2}%
\end{center}
\end{figure}

\begin{figure}
[ptb]
\begin{center}
\includegraphics[
height=2.1612in,
width=2.8738in
]%
{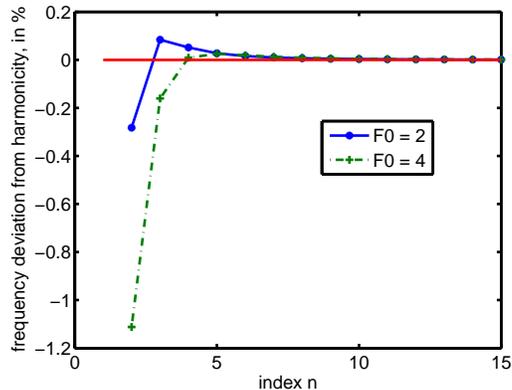}%
\caption{The deviation from harmonicity as a function of the eigenfrequency
index, for two different inhomogeneities. This deviation is defined in terms
of the difference between two neighboring frequencies $d(n)=[w(n)-w(n-1)]$ as
$\{d(n+1)/d(n)-1\}\ast100.$ The inhomogeneity is given by $R(x)=1+F_{0}%
\ x^{2}$ with $F_{0}$ either $2~$or$~4.$}%
\label{FIG9s_4}%
\end{center}
\end{figure}

The results for the eigenvalues $\Lambda_{n}$ are shown in Fig. (\ref{FIG9s_1}%
)
and the corresponding frequencies are shown in Fig. (\ref{FIG9s_2}). The
frequencies for the homogeneous string, i.e., for $R(x)=1$, are shown by the
open circles in Fig. (\ref{FIG9s_2}). Since the inhomogeneous string is more
dense at large values of $x$ than the homogeneous one, the corresponding
eigenfrequencies are correspondingly smaller. It is noteworthy that the
eigenfrequencies of the inhomogeneous string nearly fall on a straight line,
which means that the frequencies are nearly equispaced, i.e., they nearly
follow the same harmonic relationship as the ones for the homogeneous string.
The physical explanation for this property has not been investigated here, but
could be connected to the fact that the waves for the high indices have more
nodes than for the low indices, and hence lead to better averaging in a
variational procedure.%
Near the fundamental frequency slight deviations from harmonicity do occur, as
illustrated in Fig. (\ref{FIG9s_4}).%
However, small deviations from harmonicity will also be caused by other
effects such as the stiffness of the string.

\begin{figure}
[ptb]
\begin{center}
\includegraphics[
height=2.2451in,
width=2.9853in
]%
{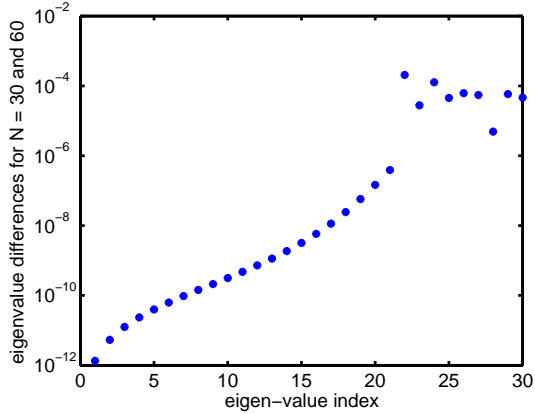}%
\caption{The dependence of the eigenvalues of the matrix $M_{fourier}$ on the
dimension $N\times N$ of the matrix. The $y-$axis shows the absolute value of
the difference between two sets of eigenvalues, one for $N=30,$ the other for
$N=60$. Some numerical values are given in Table \ref{T1}.}%
\label{FIG9s_3}%
\end{center}
\end{figure}

Figures (\ref{FIG9s_1}) and (\ref{FIG9s_2}) show that for the truncation value
$N$ of $30$, the eigenvalues become unreliable for $n\geq22$. This is a
general property of the high-n eigenvalues of a matrix, which however can be
overcome by using the iterative method described further on. The table
\ref{T1} and Fig. (\ref{FIG9s_3}) give a quantitative illustration of the
dependence of the eigenvalue on the truncation value $N$ by the comparison of
two eigenvalues for the same $n$\ of the matrix $\mathbf{M}_{fourier}%
(30\times30)$ with those of $\mathbf{M}_{fourier}(60\times60)$.
\begin{table}[tbp] \centering
\begin{tabular}
[c]{|l|l|l|}\hline
\textbf{n} & $\mathbf{N=30}$ & $\mathbf{N=60}$\\\hline\hline
\textbf{1} & 1.614775590198150e-001 & 1.6147755902115e-001\\\cline{2-3}%
\textbf{20} & 4.092e-004 & 4.0933853097811e-004\\\hline
\end{tabular}
\caption{Eigenvalues of the matrix M for two different dimensions N x N}\label{T1}%
\end{table}%
\newline%

\section{\bigskip Spectral expansions into Chebyshev Polynomials}

First some basic properties of Chebyshev polynomials will be described, then
the Curtis-Clenshaw method for expanding functions in terms of these
polynomials will be presented, with special emphasis on the errors associated
with the truncation of the expansion, and finally the application to solving
integral equations will be presented.

\subsection{Properties of Chebyshev Polynomials}

Chebyshev Polynomials $T_{v}(x)$ provide a very useful set of basis functions
for expansion purposes \cite{RUDIN}, \cite{LUKE}. A short review of the main
properties needed for the present application is presented below. The variable
$x$ is contained in the interval $-1\rightarrow+1$, and is related to an angle
$\theta$ by $x=\cos\theta$. This shows that the $x^{\prime}s$ are projections
on the $x-$axis of the tip of a radius vector of unit length that describes a
semi-circle as $\theta$ goes from $0$ to $\pi$. In terms of the $x-$variable
the $T_{n}$'s are given by
\begin{equation}%
\begin{array}
[b]{c}%
T_{0}=1\\
T_{1}=x\\
T_{2}=2x^{2}-1\\
T_{n+1}=2xT_{n}-T_{n-1}%
\end{array}
\label{6-1}%
\end{equation}
In terms of the $\theta$ variable they are given by
\begin{equation}
T_{n}=\cos(n\ \theta);~~~0\leq\theta\leq\pi. \label{6-2}%
\end{equation}

\begin{figure}
[ptb]
\begin{center}
\includegraphics[
height=2.354in,
width=3.1315in
]%
{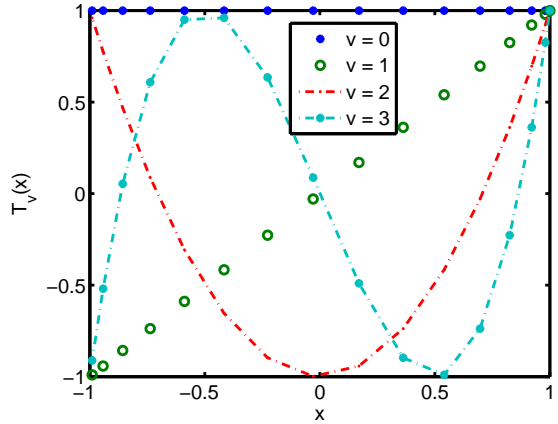}%
\caption{Chebyshev Polynomials for indices $v=0,\ 1,2,$ and $3$. They are
calculated from $T_{v}(x)=\cos(v\ast\theta),$ for the equispaced angles
$\theta.$}%
\label{FIG6_4}%
\end{center}
\end{figure}

It is clear from Eq. (\ref{6-2}) that $-1\leq T_{n}(x)\leq1$, and that the
larger the index $n$, the more zeros these polynomials have. The
$T_{n}^{\prime}s$ are orthogonal to each other with the weight function
$(1-x^{2})^{-1/2}.$ The integral $\mathcal{I}$%
\begin{equation}
\mathcal{I}_{n,m}\mathcal{=}\int_{-1}^{+1}T_{n}(x)\ T_{m}(x)\ (1-x^{2}%
)^{-1/2}\ dx=\int_{0}^{\pi}\ \cos(n\theta)\ \cos(m\theta)\ d\theta\label{6-3}%
\end{equation}
has the value $0$ if $n\neq m$, and the values $\pi/2$ if $n=m\neq0$ and $\pi$
if $n=m=0.$ A plot of $T_{v}(x)$ for $v=0,1,2,$ and $3$ is shown in Fig.
(\ref{FIG6_4}), which also illustrates that for equispaced values of $\theta$
the corresponding values of $x$ are not equispaced.%

The values of $x,$ denoted as $\xi_{i},$ for which a particular $T_{n}=0,$ are
also not equispaced. As can be seen from Eq. (\ref{6-2}) the zeros $\xi_{i}$
of $T_{N+1}(x)$ with $i=0,1,2,..N$ \ are given by
\begin{equation}
\xi_{i}=\cos\left[  \frac{(2\ i+1)}{2N+2}\pi\right]  ~\ i=0,1,2,..N.
\label{6-6b}%
\end{equation}

\subsection{The Expansion Method}

Given a function $f(r)$, defined in the interval $[a,b],$ in order to expand
it into Chebyshev polynomials, the first step is to transform the variable $r$
to a new variable $x$ defined in the interval $[-1,+1].$ This can be achieved
by means of the linear transformation
\begin{equation}
r=\alpha\ x+\beta, \label{6-1b}%
\end{equation}
with $\alpha=(b-a)/2$ and $\beta=(b+a)/2.$ In terms of the $x-$variable one
obtains the function $\bar{f}(x)=f(r),$ and the desired (truncated) expansion
is
\begin{equation}
\bar{f}^{(N)}(x)=\sum_{n=0}^{N}\ a_{n}\ T_{n}(x). \label{6-2b}%
\end{equation}
The conventional method of obtaining the expansion coefficients $a_{n}$ is to
multiply Eq. (\ref{6-2b}) on both sides by $T_{m}(x)/\sqrt{1-x^{2}}$,
integrate over $x$ from $-1$ to $+1$, and use the orthogonality condition
(\ref{6-3}). A more computer friendly alternative was given by Clenshaw and
Curtis \cite{CC}. It consists in writing Eq. (\ref{6-2b}) $N+1$ times for the
zeros $\xi_{0},\xi_{1},...\xi_{N},$ of the first Chebyshev polynomial
$T_{N+1}$ not included in the sum (\ref{6-2b}), and thus obtain $N+1$ linear
equations for the $N+1$ coefficients,
\begin{align*}
\bar{f}^{(N)}(\xi_{0})  &  =\sum_{n=0}^{N}\ a_{n}\ T_{n}(\xi_{0})\\
\bar{f}^{(N)}(\xi_{1})  &  =\sum_{n=0}^{N}\ a_{n}\ T_{n}(\xi_{1})..\\
&  \vdots\\
\bar{f}^{(N)}(\xi_{N})  &  =\sum_{n=0}^{N}\ a_{n}\ T_{n}(\xi_{N}).
\end{align*}
which in matrix notation has the form
\begin{equation}%
\begin{pmatrix}
\bar{f}(\xi_{0})\\
\bar{f}(\xi_{1})\\
\vdots\\
\bar{f}(\xi_{N})
\end{pmatrix}
=C\ast%
\begin{pmatrix}
a_{0}\\
a_{1}\\
\vdots\\
a_{n}%
\end{pmatrix}
\label{6-7b}%
\end{equation}
where $C$ is known as the Discrete Cosine Transform. The points $\xi_{i}$ are
denoted as "support points" of the algorithm since the function $\bar{f}$ has
to be known only at these points. The elements of the matrix $C$ are
$C_{i,j}=T_{j}(\xi_{i}),$ and its columns are orthogonal to each other. After
column normalization, one obtains an orthogonal matrix, and hence the inverse
$C^{-1}$ can be easily obtained, without the need to invoke a numerical matrix
inversion algorithm. The matrix $C^{-1}$ is denoted as $CM1$ in the MATLAB
program $[C,CM1,z]=C\_CM1(N)$ available in Ref. \cite{PADRE}. The row vector
$z$ contains the values $\xi_{i}$ in descending order $i=N,\ N-1,\ ..0$.
Inserting the values of $a_{i}$, obtained from Eq. (\ref{6-7b}) into Eq.
(\ref{6-2b}), one obtains the value of the truncated function $\bar{f}%
^{(N)}(x)$ at any point in the interval $[-1,+1]$, and hence the procedure is
an interpolation method \cite{DELOFF}, \cite{SPECTRAL}. Other cosine
transforms also do exist, for example one based on the Fourier series
expansion method. The method is computationally fast, in view of the advent of
the FFT algorithms, however a comparison\ of the spectral method with this
method is beyond the scope of the present article.

How good is approximation (\ref{6-2b}) to $\bar{f}(x)?$ If the function is
differentiable $p$ times, then it can be shown \cite{ORZAG} that
\begin{equation}
|\bar{f}^{(N)}(x)-\bar{f}(x)|\ \leq\frac{c}{p-1}\frac{1}{N^{p-1}} \label{6-8b}%
\end{equation}
where $c$ is a constant that depends on the $p$'s derivative of $\bar{f}.$ If
the function $\bar{f}$ is infinitely differentiable, then $p=\infty$, and the
error (\ref{6-8b}) decreases with $N$ faster than any power of $N.$ This is
denoted as the supra-algebraic convergence of the approximation of $\bar
{f}^{(N)}(x)$ to $\bar{f}(x)$, a property also denoted as "spectral" expansion
of $\bar{f}(x)$ in terms of Chebyshev polynomials \cite{ORZAG}, \cite{DELOFF}.

According to Luke \cite{LUKE}, Theorem 2 in Chapter XI, section 11.7
\begin{equation}
|\bar{f}^{(N)}(x)-\bar{f}(x)|\ \simeq a_{N+1}T_{N+1}(x)\left[  1+2x\ a_{N+1}%
/a_{N+2}\right]  \label{6-10b}%
\end{equation}
In practice,%
\begin{equation}
|\bar{f}^{(N)}(x)-\bar{f}(x)|\ \leq\ |a_{N+1}| \label{6-9b}%
\end{equation}
This property enables one to pre-assign an accuracy requirement $tol$ for the
expansion (\ref{6-2b}). Either, for a given value of $N,$ the size of the
partition of $r$ within which the function $f(r)$ is expanded can be
determined, or, for a given size of the partition, the value of $N$ can be
determined, such that the sum of the absolute values of the three last
expansion coefficients $a_{N-2},\ a_{N-1}$ and $a_{N}$ is less than the value
of $tol.$\textbf{\bigskip}

\begin{figure}
[ptb]
\begin{center}
\includegraphics[
height=2.3186in,
width=3.0822in
]%
{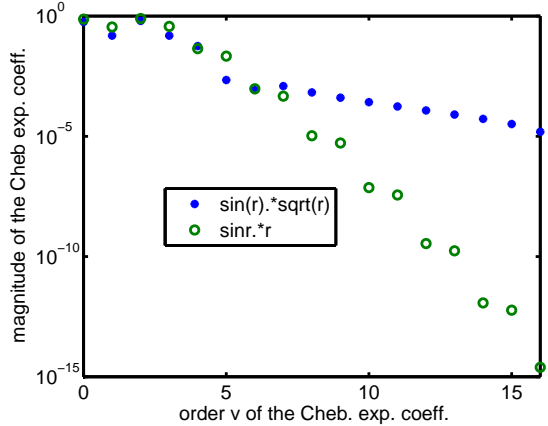}%
\caption{The Chebyshev expansion coefficients as a function of the index $v$,
for the functions $f_{1}$ and $f_{2}$ defined in Eqs. (\ref{6-3b}) and
(\ref{6-5b}). Since the derivatives of the function $f_{1}$ have a singularity
at the origin, the Chebyshev expansion converges more slowly than that of
$f_{2}$, which has an infinite number of non-singular derivatives.}%
\label{FIG6-5}%
\end{center}
\end{figure}

An example will now be given that shows that, if the function $f$ is not
infinitely differentiable, then the corresponding Chebyshev expansion
converges correspondingly slowly. The two functions to be expanded are
\begin{equation}
f_{1}(r)=r^{1/2}\sin(r) \label{6-3b}%
\end{equation}%
\begin{equation}
f_{2}(r)=r\ \sin(r) \label{6-5b}%
\end{equation}
in the interval $0\leq r\leq\pi.$ While $f_{2}$ is infinitely differentiable,
all the derivatives of the function $f_{1}$ are singular at $r=0.$ The results
for the Chebyshev expansions for the functions $f_{1}$ and $f_{2}$ using the
Clenshaw-Curtis method are displayed in Fig. (\ref{FIG6-5}).

\begin{figure}
[ptb]
\begin{center}
\includegraphics[
height=1.932in,
width=2.5685in
]%
{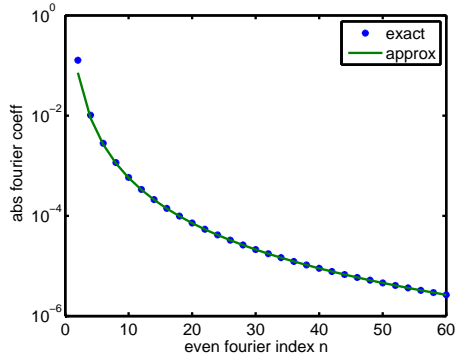}%
\caption{The Fourier expansion coefficients of the function $f(x)=x\sin
(\pi\ x)$ in the interval $[0,1]$ in terms of the basis functions $\sqrt{\pi
}\sin(n\pi\ x).$ The analytic result, given by Eqs. (\ref{II-9}) and
(\ref{II-8}) with $L=1$, is shown by the symbols $\ast$. For odd values of
$n\neq1$ they are zero. The solid line represents an approximation to
$\vert$$a_{n}|\simeq4\sqrt{\frac{2}{\pi}}(1/n)^{3}$.}%
\label{FIG6-8}%
\end{center}
\end{figure}

An expansion into a Fourier series of the function $f_{2}(r)=r\sin(\pi r)$ for
$[0\leq r\leq1]$ is also carried out for comparison with the expansion into
Chebyshev polynomials. One finds that all Fourier coefficients $a_{n}$ with
$n=1,2..$ defined in Eqs. (\ref{9-11}) through (\ref{II-8}) for $L=1,$ vanish
for $n$ odd, with the exception for $n=1$. For $n\gg1$, $a_{n}$ will approach
$0$ like $-4\sqrt{\frac{2}{\pi}}(1/n)^{3},$ i.e., quite slowly. The absolute
value of this result is shown in Fig.(\ref{FIG6-8}). By comparison with
Fig.(\ref{FIG6-5}) one sees that the Fourier expansion coefficients decrease
with the index $n$ much more slowly than the Chebyshev expansion
coefficients.

\subsection{Integrals based on spectral expansions.}

Given a function $f(r)$, defined in an interval $a\leq r\leq b$, it is the
purpose of this sub-section to numerically obtain a spectral approximation to
the indefinite integral of this function%
\begin{equation}
\mathfrak{I(}r\mathfrak{)}=\int_{a}^{r}f(r^{\prime})dr^{\prime} \label{6-1c}%
\end{equation}
As is done in Eq. (\ref{6-1b}) the function $f(r)$ is transformed from the
variable $r$ to the function $\bar{f}(x)$ for the variable $x\subset
\lbrack-1,+1]$. Then the integral (\ref{6-1c}) becomes%
\begin{equation}
\mathfrak{I(}r\mathfrak{)=}\frac{\mathfrak{(}b-a)}{2}I_{L}(x) \label{6-2c}%
\end{equation}
where%
\begin{equation}
I_{L}(x)=\int_{-1}^{x}\bar{f}(x^{\prime})dx^{\prime}. \label{6-3c}%
\end{equation}
It is desired to obtain the spectral expansion of the approximation to
$I_{L}(x)$
\begin{equation}
I_{L}^{(N)}(x)=\sum_{n=0}^{N}\ b_{n}\ T_{n}(x) \label{6-4c}%
\end{equation}
where it is assumed that $\bar{f}(x)$ has been expanded in a series of
Chebyshev polynomials, as given by Eq. (\ref{6-2b}). In view of the integral
properties of Chebyshev polynomials, the coefficients $b_{n}$ can be expressed
in terms of the expansion coefficients $a_{n}$ of $\bar{f}(x),$%
\begin{equation}%
\begin{pmatrix}
b_{0}\\
b_{1}\\
\vdots\\
b_{N}%
\end{pmatrix}
=S_{L}%
\begin{pmatrix}
a_{0}\\
a_{1}\\
\vdots\\
a_{N}%
\end{pmatrix}
\label{6-5c}%
\end{equation}
by means of the matrix $S_{L}$ \cite{CC}, without loss of accuracy. For the
integral%
\begin{equation}
I_{R}(x)=\int_{x}^{1}\bar{f}(x^{\prime})dx^{\prime} \label{6-6c}%
\end{equation}
an expression similar to (\ref{6-5c}) exists, with the matrix $S_{L}$ replaced
by $S_{R}.$ Numerical expressions for the matrices $S_{L}$ and $S_{R}$ exist
in the literature \cite{IEM}, \cite{DELOFF-INT} and are also available from
Ref. \cite{PADRE} under the name $SL\_SR.$ In particular, by noting that
$T_{n}(1)=1$ for all $n,$ an approximation to the definite integral
$\mathfrak{I(}r_{2}\mathfrak{)}=\int_{a}^{b}f(r^{\prime})dr^{\prime}$ is given
by%
\begin{equation}
\mathfrak{I}^{(N)}\mathfrak{(}b\mathfrak{)=}\frac{\mathfrak{(}a-b)}{2}%
\sum_{n=0}^{N}b_{n} \label{6-8c}%
\end{equation}
with an error comparable to Eq. (\ref{6-9b}), of the order of $|b_{N+1}|.$ The
above form of the definite integral (\ref{6-8c}) is denoted below as
Gauss-Chebyshev quadrature. The existence of Eq. (\ref{6-5c}) makes the
expansion into Chebyshev polynomials very suitable for the numerical solution
of integral equations, as will be seen below.

As an example, the integrals
\begin{align}
\mathfrak{I}_{1}  &  =\int_{0}^{\pi}r^{1/2}\ \sin(r)\ dr\label{4b}\\
\mathfrak{I}_{2}  &  =\int_{0}^{\pi}r\ \sin(r)\ dr \label{5b}%
\end{align}
are evaluated below by using Eq. (\ref{6-8c}). For comparison purposes
$\mathfrak{I}_{1}$ was also evaluated using the MATLAB integration function
$quad(@myfun,0,\pi,acc),$ where $acc$ denotes the precision to within which
the quadrature result is given. The results \ are shown in the last line of
table \ref{TABLE4}

\begin{figure}
[ptb]
\begin{center}
\includegraphics[
height=2.3333in,
width=3.103in
]%
{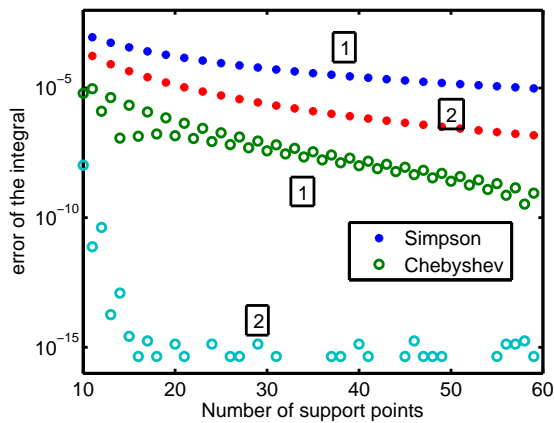}%
\caption{Comparison of the convergence properties of the Gauss-Chebyshev and
the Simpson integration procedures as a function of the number of support
points. The labels $1$ or $2$ denote the integrals $I_{1}=\int_{0}^{\pi}%
\sin(r)\ r^{1/2}\ dr$ or \ $I_{2}=\int_{0}^{\pi}\sin(r)\ r\ dr,$
respectively.}%
\label{FIG6-7}%
\end{center}
\end{figure}

If one uses an expansion of the integrand $f(r)=\sin(r)\ast r^{1/2}$ into a
set of Chebyshev polynomials, and uses the integral properties of these
polynomials by means of the function $[SL,SR]=SL\_SR(N)$, then for $60$
support points $N$ one gets an accuracy of $1:10^{-9},$ but the convergence
with $N$ is slow, as is also the case for the expansion coefficients of
$f_{1}$. The values of $\mathfrak{I}_{1}$ for two values of $N$ are shown in
Table \ref{TABLE4} below%
\begin{table}[tbp] \centering
\begin{tabular}
[c]{|l|l|}\hline
$Cheb.,\ N=58$ & $2.43532116647$\\\hline
$Cheb.,\ N=59$ & $2.43532116702$\\\hline
$quad,\ acc=10^{-11}$ & $2.4353211641\mathbf{7}$\\\hline
\end{tabular}
\caption{Integral (\ref{4b}) obtained with the Chebyshev method, for benchmark purposes}\label{TABLE4}%
\end{table}%
.\ If, on the other hand, one instead uses for the integrand the analytic
function $f_{2}(r)=\sin(r)\ast r,$ then the corresponding integral converges
with $N$ much faster, reaching machine accuracy for $N=18.$ These convergence
properties are displayed in Fig. (\ref{FIG6-7}), where a comparison of the
convergence using Simpson's quadrature method is also shown.%

\section{The integral equation for the inhomogeneous string.}

In the previous discussion the Sturm-Liouville functions $\psi_{n}(r),$
solutions of Eq. (\ref{9-5}), were obtained by expanding them into a set of
Fourier functions $\phi_{\ell}(r),$ and obtaining the eigenfunctions and
eigenvalues of the matrix $M_{fourier}.$ This matrix consisted of overlap
integrals of the inhomogeneity function $R(x)$ with the basis functions
$\phi_{\ell}(x).$ In the present section three major innovations are
introduced:\ a) we transform the differential equation (\ref{9-5}) into an
integral equation, since the numerical solution of the latter is more stable
than that of the former, b) we replace the need to do overlap integrals by the
Curtis Clenshaw method, Eq. (\ref{6-7b}), of obtaining the expansion
coefficients, and c) the basis functions are the Chebyshev polynomials for
which the expansion series converges much faster than for the Fourier expansions.

The integral equation that is equivalent to the differential equation
(\ref{9-5}) is
\begin{equation}
\frac{1}{\Lambda}\psi(r)=-\int_{0}^{L}\mathcal{G}(r,r^{\prime})\ R(r^{\prime
})\ \psi(r^{\prime})dr^{\prime} \label{10_2}%
\end{equation}
where the Green's function $\mathcal{G}(r,r^{\prime})$ is given by%
\begin{align}
\mathcal{G}(r,r^{\prime})  &  =-\frac{1}{L}F(r)G(r^{\prime})\text{ for
}r<r^{\prime}\label{10_3}\\
\mathcal{G}(r,r^{\prime})  &  =-\frac{1}{L}F(r^{\prime})G(r)\text{ for
}r>r^{\prime}\nonumber
\end{align}
and where
\begin{equation}
F(r)=r;~~G(r)=(L-r). \label{10_4}%
\end{equation}
Both functions $F$ and $G$ obey the equation $d^{2}F/dr^{2}=0,$ $d^{2}%
G/dr^{2}=0$\ and they are linearly independent of each other. Because of the
separable nature of $\mathcal{G}$ the integral on the right hand side of Eq.
(\ref{10_2}) can be written as
\begin{align}
\int_{0}^{L}\mathcal{G}(r,r^{\prime})\ R(r^{\prime})\ \psi(r^{\prime
})dr^{\prime}  &  =-\frac{1}{L}G(r)\int_{0}^{r}F(r^{\prime})R(r^{\prime
})\ \psi(r^{\prime})dr^{\prime}\nonumber\\
&  -\frac{1}{L}F(r)\int_{r}^{L}G(r^{\prime})R(r^{\prime})\ \psi(r^{\prime
})dr^{\prime} \label{10_5}%
\end{align}

In view of the fact that $F$ vanishes at $r=0$ and $G$ vanishes at $r=L,$ and
hence $\int_{0}^{L}\mathcal{G}(r,r^{\prime})\ R(r^{\prime})\ \psi(r^{\prime
})dr^{\prime}$ vanishes for both $r=0$ and $r=L,$ the functions $\psi$ satisfy
the boundary conditions$.$ A proof that $\psi(r)$ defined by Eq. (\ref{10_2})
satisfies Eq. (\ref{9-5}) can be obtained by carrying out the second
derivative in $r$ of Eq. (\ref{10_5}).

The numerical solution of Eq. (\ref{10_2}) is accomplished by first changing
the variable $r$, contained in the interval $[0,L]$, into the variable $x$,
contained in the interval $[-1,+1],$\ which results in the transformed
functions $\bar{\psi}(x)$, $\mathcal{\bar{G}}(x,x^{\prime})$, and $\bar
{R}(x^{\prime}).$ Expanding the unknown solution $\bar{\psi}(x)$ into
Chebyshev polynomials%
\begin{equation}
\bar{\psi}(x)=\sum_{n=0}^{N}\ a_{n}T_{n}(x), \label{10_6}%
\end{equation}
as was done in Eq. (\ref{6-2b}), then Eq. (\ref{10_2}) leads to a matrix
equation in the coefficients $a_{n}$, as will now be shown. The coefficients
$a_{n}$ can be placed into a column vector%
\begin{equation}
\vec{a}=[a_{0},a_{1},..,a_{N}]^{T}, \label{10_7}%
\end{equation}
where $T$ means transposition. The values of $\bar{\psi}(\xi_{i})$ at the
support points $\xi_{i}$, which are the zeros of $T_{N+1},$ can also be
expressed as a column vector%
\begin{equation}
\vec{\psi}=[\bar{\psi}(\xi_{0}),\ \bar{\psi}(\xi_{1}),\ ...\bar{\psi}(\xi
_{N})]^{T}, \label{10_8}%
\end{equation}
and the relation between $\vec{a}$ and $\vec{\psi},$ already given in Eq.
(\ref{6-7b}), is%
\begin{equation}
\vec{a}=\mathbf{C}^{-1}\ \vec{\psi},~~\vec{\psi}=\mathbf{C\ }\vec{a}
\label{10_9}%
\end{equation}

Another important relation concerns the integrals%
\begin{equation}
\Phi_{L}(x)=\int_{-1}^{x}\phi(x^{\prime})\ dx^{\prime}\text{ and }\Phi
_{R}(x)=\int_{x}^{1}\phi(x^{\prime})\ dx^{\prime}, \label{10_10}%
\end{equation}
where $\phi$ is a function defined in the interval $[-1,1]$, and the
corresponding expansion coefficients $\alpha_{n}$ are given by $\vec{\alpha
}=\mathbf{C}^{-1}\ \vec{\phi}$. If $\Phi_{L,R}(x)$ is expanded into Chebyshev
polynomials%
\begin{equation}
\Phi_{L}(x)=\sum_{n=0}^{n=N}\ \beta_{n}^{(L)}T_{n}(x)\text{ and~ }\Phi
_{R}(x)=\sum_{n=0}^{n=N}\ \beta_{n}^{(R)}T_{n}(x) \label{10_11}%
\end{equation}
then the expansion coefficients $\beta$ can be expressed in terms of the
expansion coefficients $\alpha$ of $\phi$ by means of the matrices
$\mathbf{S}_{L}$ and $\mathbf{S}_{R},$ described near Eq. (\ref{6-5c}),%
\begin{equation}
\vec{\beta}^{(L)}=\mathbf{S}_{L}\vec{\alpha}\text{ and }\vec{\beta}%
^{(R)}=\mathbf{S}_{R}\vec{\alpha}\text{.} \label{10_12}%
\end{equation}
The matrices $\mathbf{C,\ C}^{-1},\ \mathbf{S}_{L}$ and $\mathbf{S}_{R}$ can
either be obtained from Ref. \cite{PADRE} or can be found in Ref.\cite{IEM}.
Making use of Eqs. (\ref{10_9}) and (\ref{10_12}) one can write the Chebyshev
expansion of the right and left hand sides of Eq. (\ref{10_2}) as%
\begin{equation}
\frac{1}{\Lambda}\vec{a}=\mathbf{M}_{IEM}\ \vec{a} \label{10_13}%
\end{equation}
where
\begin{equation}
\mathbf{M}_{IEM}=\frac{1}{2}\ast\mathbf{C}^{-1}\ast M3\ast DR\ast\mathbf{C}.
\label{10_14}%
\end{equation}
In the above the factor $1/2$, comes from the transformation of coordinates
from $r$ to $x,$ and where the term $L$ was cancelled by the $(1/L)$ in Eq.
(\ref{10_5}); $DR$ is the diagonal matrix that contains the values of
$R(\xi_{i})$ along the main diagonal, and $M_{3}$ is given by%
\begin{equation}
\mathbf{M}_{3}=DG\ast\mathbf{C}\ast S_{L}\ast\mathbf{C}^{-1}\ast
DF+DF\ast\mathbf{C}\ast S_{R}\ast\mathbf{C}^{-1}\ast DG. \label{10_15}%
\end{equation}
The first (second) term in Eq. (\ref{10_15}) represents the first (second)
term in Eq. (\ref{10_5}), $DF=diag(F)$ and\ $DG=diag(G)$ represent the
diagonal matrices having the values of $F(\xi_{i})$ and $G(\xi_{i})$ along the
main diagonal, the $\xi_{i}$ being the $N+1$ support points described near Eq.
(\ref{6-7b}).

The explanation for Eq. (\ref{10_13}) is as follows: the matrix $\mathbf{M}%
_{IEM}$ in Eq. (\ref{10_14}) is applied to the column vector $\vec{a},$ the
$\mathbf{C}$ in (\ref{10_14}) transforms the $\vec{a}$ into the vector
$\vec{\psi},$ the factor $DR$ together with the factor $DG$ in (\ref{10_15})
transforms $\vec{\psi}$ into $\vec{G}\otimes\vec{R}\otimes\vec{\psi}$ (the
symbol $\otimes$ means that in $\vec{G}\otimes\vec{R}$ each element of the
vector $\vec{G}$ is multiplied by the corresponding element of the vector
$\vec{R}$, and a new vector of the same length is produced), the additional
factor $\mathbf{C}^{-1}$ produces the expansion coefficients of $\vec
{G}\otimes\vec{R}\otimes\vec{\psi}$, the matrix $S_{L}$ or $S_{R}$ transforms
these expansion coefficients to the expansion coefficients of the respective
indefinite integrals, etc.

\subsection{Results}

After choosing a certain value for the number $N_{IEM}+1$ of Chebyshev
coefficients a numerical value of the ($N_{IEM}+1)\times(N_{IEM}+1)$ matrix
(\ref{10_14})\ is obtained, from which the eigenvalues $(1/\Lambda_{n})$,
$n=1,2,...N_{IEM}+1$ can be calculated. \ The MATLAB computing times for the
Fourier method for $N_{fourier}=30$ and $60$ combined using the analytic
expressions for the integrals needed to obtain the elements of the matrix
$\mathbf{R}$ is $0.91s$, while the computing time for the $IEM$ matrix method
for all three $N_{IEM}=30,\ 60,$ and $90$ values combined is $0.75s.$ Hence
the IEM method is comparable in complexity to the Fourier expansion
method,\ provided that the overlap integrals (\ref{9-14}) are known
analytically. However, a disadvantage of the $IEM$ for this application is
that some eigenvalues are spurious. Their occurrence can be recognized in that
they change with the value of $N_{IEM},$ and do not coincide with the
eigenvalues of $\mathbf{M}_{fourier}.$

\begin{figure}
[ptb]
\begin{center}
\includegraphics[
height=2.2667in,
width=3.013in
]%
{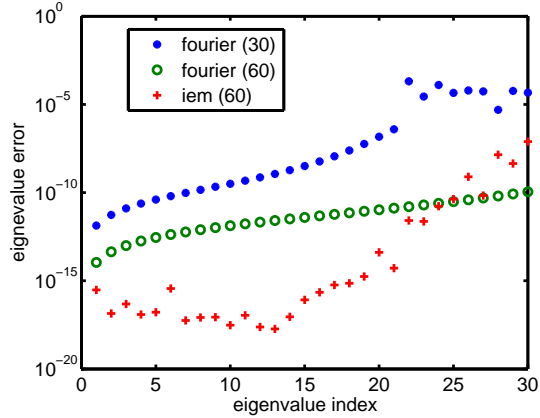}%
\caption{Accuracy of the eigenvalues of $M_{fourier}$ and $M_{IEM}$ for
various values of their dimension $N\times N.$ The value of $N$ is indicated
in parenthesis in the legend. For the Fourier method, $N$ is the number of
basis functions $\phi_{\ell}$ used to expand the Sturm-Liuoville
eigenfunctions, and for the $IEM$, $N$ is the number of Chebyshev polynomials
used in the expansion, which is also equal to the number of support points in
the interval $[0,L].$ The accuracy of the matrix eigenvalues is obtained by
comparison with a highly accurate result of $1$ part in $10^{11}$ obtained by
an iterative method. }%
\label{FIG10_3}%
\end{center}
\end{figure}

The accuracy of these two matrix methods is illustrated in Fig. (\ref{FIG10_3}%
). It is based on the iterative method described below, used as an accuracy
benchmark, since it gives an accuracy of $1:10^{11}$ for the eigenvalues
regardless of the value of the eigenvalue index $n.$
Figure (\ref{FIG10_3}) shows that the accuracy of the IEM matrix method is
considerably higher than the Fourier matrix method for the low values of $n$,
but it is not as monotonic as the latter. The figure also shows that the
accuracy of both matrix methods depends sensitively on the dimension $N$ of
their respective matrices $M$.

\section{The iterative method}

This iterative method was introduced by Hartree \cite{HART} in the 1950's in
order to calculate energy eigenvalues of the Schr\"{o}dinger equation for
atomic systems. The method was adapted to the use of the spectral expansion
method ($IEM$) and applied to the energy eigenvalue of the very tenuously
bound Helium-Helium dimer \cite{HEHE}. The version described below for finding
the eigenvalues that multiply the inhomogeneity function $R,$ with appropriate
modifications is also suitable for finding the eigenfunctions for more general
SL equations, such as the Schr\"{o}dinger equation \cite{STURM}. The method is
as follows.

For a slightly wrong value $\Lambda_{1}$ of $\Lambda$\ there is a slightly
wrong function $\psi_{1}$ that obeys the equations%
\begin{equation}
\frac{d^{2}\psi_{1}(r)}{dr^{2}}+\Lambda_{1}R(r)\ \psi_{1}(r)=0. \label{10b_3}%
\end{equation}
This function does not satisfy the boundary conditions at both $r=0$ and $r=L$
unless it has a discontinuity at some point $r_{I}$, contained in the interval
$[0,L].$ To the left of $r_{I}$ the function $\psi_{1}$ that vanishes at $r=0$
is called $Y_{1}(r),$ and to the right of $r_{I}$ it is called $k\ast
Z_{1}(r),$ and vanishes at $r=L.$ Here $\mathfrak{k}$ is a normalization
factor chosen such that $Y_{1}(r_{I})=\mathfrak{k}$\ $Z_{1}(r_{I}).$ Both
these functions rigorously obey Eq. (\ref{10b_3}) in their respective
intervals and are obtained by solving the integral equations
\begin{equation}
Y_{1}(r)=F(r)-\Lambda_{1}\int_{0}^{r_{I}}\mathcal{G}(r,r^{\prime})R(r^{\prime
})Y_{1}(r^{\prime})dr^{\prime},~~0\leq r\leq r_{I} \label{10b_4}%
\end{equation}
and%
\begin{equation}
Z_{1}(r)=G(r)-\Lambda_{1}\int_{r_{I}}^{L}\mathcal{G}(r,r^{\prime})R(r^{\prime
})Z_{1}(r^{\prime})dr^{\prime},~~r_{I}\leq r\leq L. \label{10b_5}%
\end{equation}
These integral equations differ from Eq. (\ref{10_2}), due to the presence of
a driving term $F$ or $G$. However, since the second derivatives of these
functions are zero, their presence does not prevent that $Y_{1}$ and $Z_{1}$
obey Eq. (\ref{10b_3}) in their respective domains.

The iteration from $\Lambda_{1}$ to a value closer to the true $\Lambda$
proceeds as follows. One multiplies Eq. (\ref{10b_6})
\begin{equation}
\frac{d^{2}Y_{1}(r)}{dr^{2}}+\Lambda_{1}R(r)\ Y_{1}(r)=0,\ 0\leq r\leq r_{I}
\label{10b_6}%
\end{equation}
with $\psi(r)$ and one multiplies Eq. (\ref{9-5}) with $Y_{1}(r),$ subtracts
one from the other, and integrates from $r=0$ to $r=r_{I}.$ One finds that
$\int_{0}^{r_{I}}(Y_{1}^{\prime\prime}\psi-\psi^{\prime\prime}Y_{1}%
)dr^{\prime}=$ $(Y_{1}^{\prime}\psi-\psi^{\prime}Y_{1})_{r_{I}}=(\Lambda
-\Lambda_{1})\int_{0}^{r_{I}}Y_{1}\psi dr^{\prime}$. Here a prime denotes the
derivative with respect to $r.$ A similar procedure applied to $Z_{1}$ in the
interval $[r_{I},L]$ yields $-\mathfrak{k\ }(Z_{1}^{\prime}\psi-\psi^{\prime
}Z_{1})_{r_{I}}=(\Lambda-\Lambda_{1})\int_{0}^{r_{I}}\mathfrak{k\ }Z_{1}\psi
dr^{\prime}.$ Adding these two results and remembering that $\mathfrak{k}%
Z_{1}=Y_{1}$ for $r=r_{I}$, and dividing the result by $\psi(r_{I}%
)\mathfrak{k}Z_{1}(r_{I})$ one obtains%
\begin{equation}
\Lambda-\Lambda_{1}=\frac{(Y^{\prime}/Y-Z^{\prime}/Z)_{r_{I}}}{\frac{1}%
{(Y_{1}\psi)_{r_{I}}}\int_{0}^{r_{I}}Y_{1}R\psi dr^{\prime}+\frac{1}%
{(Z_{1}\psi)_{r_{I}}}\int_{r_{i}}^{L}Z_{1}R\psi dr^{\prime}}. \label{10b_7}%
\end{equation}
This result is still exact, but the exact function $\psi$ is not known. The
iterative approximation occurs by replacing $\psi$ in the first integral in
the denominator by $Y_{1}$, and by $\mathfrak{k\ }Z_{1}$ in the second
integral, and by replacing $\psi(r_{I})$ in the denominators of each integral
by either $Y_{1}(r_{I})$ or by $\mathfrak{k\ }Z_{1}(r_{I}).$ The final result
is
\begin{equation}
\Lambda_{2}=\Lambda_{1}+\frac{(Y^{\prime}/Y-Z^{\prime}/Z)_{r_{I}}}{\frac
{1}{Y_{1}^{2}(r_{I})}\int_{0}^{r_{I}}Y_{1}^{2}Rdr^{\prime}+\frac{1}{Z_{1}%
^{2}(r_{I})}\int_{r_{i}}^{L}Z_{1}^{2}Rdr^{\prime}}. \label{10b_8}%
\end{equation}
In the above, $\Lambda$ was replaced by $\Lambda_{2}$ as being a better
approximation to $\Lambda$ than $\Lambda_{1},$ and the normalization factor
$\mathfrak{k}$ has cancelled itself out. The iteration proceeds by replacing
$\Lambda_{1}$ in the above equations by the new value $\Lambda_{2}.$

The derivatives in the numerator of Eq. (\ref{10b_8}) can be obtained without
loss of accuracy by making use of the derivatives of Eqs. (\ref{10b_4}) and
(\ref{10b_5})%
\begin{equation}
\ Y_{1}^{\prime}(r)=F^{\prime}(r)+\frac{\Lambda_{1}}{L}G^{\prime}(r)\int
_{0}^{r}F(r^{\prime})R(r^{\prime})\ Y_{1}(r^{\prime})dr^{\prime}+\frac
{\Lambda_{1}}{L}F^{\prime}(r)\int_{r}^{r_{I}}G(r^{\prime})R(r^{\prime}%
)\ Y_{1}(r^{\prime})dr^{\prime} \label{10b_9}%
\end{equation}
and%
\begin{equation}
Z_{1}^{\prime}(r)=G^{\prime}(r)+\frac{\Lambda_{1}}{L}G^{\prime}(r)\int_{r_{I}%
}^{r}F(r^{\prime})R(r^{\prime})\ Z_{1}(r^{\prime})dr^{\prime}+\frac
{\Lambda_{1}}{L}F^{\prime}(r)\int_{r}^{L}G(r^{\prime})R(r^{\prime}%
)\ Z_{1}(r^{\prime})dr^{\prime} \label{10b_10}%
\end{equation}
with the result at $r=r_{I}$%
\begin{equation}
Y_{1}^{\prime}(r_{I})=1-\frac{\Lambda_{1}}{L}\int_{0}^{r_{I}}r^{\prime
}R(r^{\prime})\ Y_{1}(r^{\prime})dr^{\prime} \label{10b_11}%
\end{equation}
and%
\begin{equation}
Z_{1}^{\prime}(r_{I})=-1+\frac{\Lambda_{1}}{L}\int_{r_{I}}^{L}(L-r^{\prime
})R(r^{\prime})\ Z_{1}(r^{\prime})dr^{\prime} \label{10b_12}%
\end{equation}

In the present formulation the dimensions of $\Lambda$ are $\ell^{-2},$ and
the dimension of $F$, $G,$ $Y$ and $Z$ are $\ell,$ where $\ell$ represents a
unit of length. As noted above, the derivatives with respect to $r$ of the
functions $Y$ or $Z$ or $\psi$ are not obtained as the difference between two
adjoining positions, but rather as the known derivatives of $F$ and $G$,
together with integrals over $Y$ or $Z$ or $\psi$ according to Eqs.
(\ref{10b_9}) and (\ref{10b_10}). In the $IEM$ formulation these integrals can
be obtained with the same spectral precision as the calculation of the
functions $Y$ or $Z$ or $\psi,$ \cite{CISE}, hence there is no loss of
accuracy either for the evaluation of Eq. (\ref{10b_8}), or for the
calculation of $\Lambda,$ which can be set to $1:10^{11}$. However, it is
important to start the iteration with a guessed value of $\Lambda$ that lies
within the valley of convergence of Eq. (\ref{10b_8}). These initial values
can be obtained, for example, from the eigenvalues of the matrix
$\mathbf{M}_{fourier}$ described above, or from a method described in Ref.
\cite{HEHE}.\bigskip

\subsection{Results for the iterative method}

Some of the values for $\Lambda_{n}$ obtained to an accuracy of $1:10^{11}$ by
means of the iterative method described above are listed in Table
\ref{Table_3}, so as to serve as benchmark results for comparisons with future
methods. The starting values $\Lambda_{1}$ for each $n$ are the results of the
Fourier method described above with $N=60.$ The iterations were stopped when
the change $\Lambda_{2}-\Lambda_{1}$ became less than $10^{-12}$ (usually
three iterations were required), and $tol=10^{-11}.$
\begin{table}[tbp] \centering
\begin{tabular}
[c]{|l|l||l|l|}\hline
$\mathbf{n}$ & $\mathbf{\Lambda}_{n}$ & $\mathbf{n}$ & $\mathbf{\Lambda}_{n}%
$\\\hline\hline
$1$ & 1.61477559021e-001 & $26$ & 2.42220326385e-004\\\hline
$2$ & 4.06257259855e-002 & $27$ & 2.24611142229e-004\\\hline
$3$ & 1.81281029690e-002 & $28$ & 2.08854647313e-004\\\hline
$4$ & 1.02131986136e-002 & $29$ & 1.94699775697e-004\\\hline
$5$ & 6.54130338213e-003 & $30$ & 1.81936592475e-004\\\hline
\end{tabular}
\caption{Eigenvalues of Eq.(\ref{9-5}) obtained iteratively with Eqs.(\ref{10b_8}) }\label{Table_3}%
\end{table}%
\bigskip

\begin{figure}
[ptb]
\begin{center}
\includegraphics[
height=2.2035in,
width=2.93in
]%
{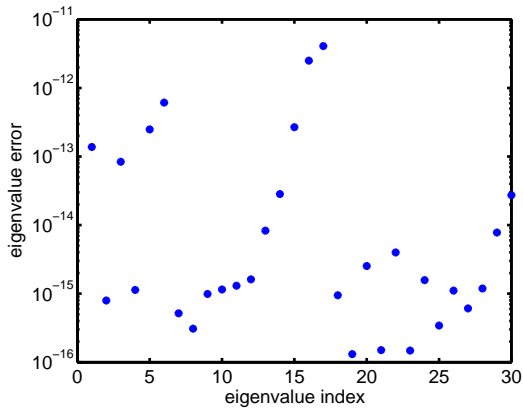}%
\caption{The $y$-axis shows the absolute value of the mean square average of
the three last Chebyshev coefficents\ \ in the expansions of the functions $Y$
and $Z.$ As discussed in the text, the error of the eigenvalues $\Lambda$ is
also given by the $y-$axis. The number $N$ of expansion Chebyshev polynomials
was increased adaptively as the eigenvalue index $n$ increased. The "jumps" in
the values of these errors is due to the transition from one value of $N$ to a
suddenly larger value, as is explained in the text.}%
\label{FIG10_4}%
\end{center}
\end{figure}

The error of the functions $Y$ and $Z$ is given, according to Eq.
(\ref{6-9b}), by the size of the high order Chebyshev expansion parameters.
For the $tol$ parameter of $10^{-11}$ their values stay below $10^{-11},$ as
is shown in Fig. (\ref{FIG10_4}). Since there is no loss of accuracy in
evaluating the various terms in Eq. (\ref{10b_8}), the error in the iterated
eigenvalues $\Lambda$ is also given by Fig. (\ref{FIG10_4}). In order to
achieve this type of error, the number $N$ of Chebyshev polynomials used for
the spectral expansion of the functions $Y$ and $Z$ for the solution of their
respective integral equations was increased adaptively by the computer
program. It was found that for $n=1$, $N=16$; for $n=2$ to $6$, $N=24$; for
$n=7$ to $23$, $N=24$; and for $n=18$ to $30$, $N=54$. This procedure of
increasing $N$ is different from the procedure used in Ref. \cite{HEHE}, where
$N$ was kept constant and the number of partitions was increased adaptively.
The latter method was required because of the long range ($3000$ units of
length) of the $He-He$ wave functions.%

\section{Summary and conclusions}

The main aim of this paper is to introduce the spectral expansion method for
solving integral equations to the teaching community, with the hope that this
method can be included in computational physics courses in the future. Such
expansions converge rapidly with high precision, and complement the usual
finite difference methods in common use today. The example used for the
application of such a method is the analysis of the vibration of an
inhomogeneous string in the separation of variables formalism. The spatial
basis functions $\psi_{n}(r),$ $n=1,2,...,$ form a complete Sturm-Liouville
set, the calculation of which is performed by means of three methods. In
method $1$ the function $\psi$\ is expanded into a basis set of sine waves,
and the eigenfrequencies and expansion coefficients for each $\psi_{n}$ are
the eigenvalues and eigenvectors of a matrix $\mathbf{M}_{fourier}.$ In method
$2$ the differential equation for $\psi_{n}$ is transformed into an integral
equation of the Lippmann Schwinger type, the unknown function is expanded into
Chebyshev polynomials, and the expansion coefficients are again the
eigenvectors of another matrix $\mathbf{M}_{IEM}.$ The comparison between
these two methods illustrates the differences and advantages of each,
especially their properties as a function of the size of the expansion basis.
In method $3$, which has not been presented previously, the differential
equation for the Sturm-Liouville eigenfunction is solved iteratively, and the
auxiliary functions required for the iterations are obtained from the
solutions of the corresponding integral equation. The advantage of method $3$
is that the precision of both the eigenfunction and the eigenvalue can be
predetermined by specifying the value of a tolerance parameter, and further,
no eigenvalue calculation of big matrices is required. In the present
application the results of method $3$ were accurate to $1:10^{11}.$

The comparison between the accuracies of the various methods is illustrated
extensively by means of appropriate graphs and tables. Applications of these
methods to other problems, such as the solution of the Schr\"{o}dinger
equation, or the heat propagation equation, or diffusion equations in biology,
are of course quite possible in spite of the present focus on the
inhomogeneous string equation.

\end{document}